\documentstyle{mn-1.4}
\newif\ifAMStwofonts
\AMStwofontstrue

\def\sun{\ifmmode\odot\else$\odot$\fi}
\def\HI{\hbox{H\,{\sc i}}}
\def\H2{\hbox{H$_{2}$}}
\def\HII{\hbox{H\,{\sc ii}}}

\def\Halpha{\hbox{\rm H$\alpha$}}

\def\Bgamma{\hbox{\rm Br$\gamma$}}
\def\Palpha{\hbox{\rm Pa$\alpha$}}
\def\COsp{\hbox{CO$_{\rm sp}$}}
\def\mic{$\mu$m}
\def\kms{${\rm km~s}^{-1}$}

\title[Circumnuclear star formation in M100]
       {Near-infrared spectroscopy of the
        circumnuclear star formation regions in M100: Evidence
        for sequential triggering}
\author[S.~Ryder, J.~Knapen and M.~Takamiya]
       {S. D. Ryder$^{1}$\footnotemark,
        J. H. Knapen$^{2}$\footnotemark, and M. Takamiya$^{3}$\\
        $^{1}$ Anglo-Australian Observatory, P.O. Box 296, Epping, NSW~1710, 
        Australia.\\
        $^{2}$ Isaac Newton Group of Telescopes, Apartado 321, E-38700
        Santa Cruz de La Palma, Spain.\\
        $^{3}$ Gemini Observatory, 670 N. A'Ohoku Place, Hilo, HI 96720,
        U.S.A.}
        \date{Accepted 21 November 2000.
      Received 6 November 2000;
      in original form 28 August 2000}

\pagerange{\pageref{firstpage}--18}
\pubyear{1999}

\begin{document}

\maketitle

\label{firstpage}

\begin{abstract}

We present low resolution ($R\sim450$) $K$-band spectroscopy for 16 of
the 43 circumnuclear star-forming knots in M100 identified by Ryder \&
Knapen (1999). We compare our measurements of equivalent widths for
the \Bgamma\ emission line and CO 2.29~\mic\ absorption band in each
knot with the predictions of starburst models from the literature, and
derive ages and burst parameters for the knots. The majority of these
knots are best explained by the result of short, localised bursts of
star formation between 8 and 10~Myr ago. By examining both radial and
azimuthal trends in the age distribution, we present a case for
sequential triggering of star formation, most likely due to the action
of a large-scale shock. In an appendix, we draw attention to the fact
that the growth in the CO spectroscopic index with decreasing
temperature in supergiant stars is not as regular as is commonly
assumed.

\end{abstract}

\begin{keywords}
galaxies: evolution -- galaxies: individual (M100, NGC 4321) --
galaxies: kinematics and dynamics -- galaxies: structure -- infrared:
galaxies.
\end{keywords}

\section[]{Introduction}

Nowhere is the interplay between galaxy dynamics and star formation
more important than in the circumnuclear regions (CNRs) of barred
spiral galaxies. Where star formation is observed to occur, it is
frequently organized into a tightly-wound spiral, or even a complete
ring, on the scale of a kiloparsec or more. Based on numerical
simulations of gas flows in a barred potential, it is thought that
these features are associated with dynamical resonances, specifically
the Inner Lindblad Resonances (ILRs). The resulting concentration of
gas in a narrow region leads eventually to the triggering of star
formation in a host of individual `hot-spots' or `knots'. However much
of the detailed specifics, such as whether star formation commences in
all such knots at the same time, and in the same manner, remain a
mystery.

\addtocounter{footnote}{-1}
\footnotetext{E-mail: sdr@aaoepp.aao.gov.au}
\addtocounter{footnote}{1}
\footnotetext{On leave from: Department of Physical Sciences,
University of Hertfordshire, Hatfield, Herts AL10 9AB.}

In an effort to address such issues, we recently obtained (Ryder \&
Knapen 1999, hereafter Paper I) high-resolution images in the $J$,
$H$, and $K$ bands with IRCAM3 on the United Kingdom Infrared
Telescope (UKIRT) of the circumnuclear region of M100 (NGC~4321), the
brightest spiral galaxy in the Virgo cluster. In conditions of
$<0.4$~arcsec seeing, we were able to identify 43~compact knots in the
$K$ image, and to determine magnitudes and colours for 41 of
these. Spectroscopic observations with CGS4 on UKIRT, with the slit
position angle set to cross the nucleus and the two large-scale
hot-spots K1 and K2 identified previously by Knapen et al. (1995a),
revealed the presence of weak \Bgamma\ and \H2\ 1--0 S(1) emission
lines, as well as deep CO (2--0) absorption bands in both K1 and K2. A
comparison with models for starburst galaxies by Puxley, Doyon \& Ward
(1997; hereafter PDW) indicated that these knots were formed between
15 and 25~Myr ago in a burst of star formation which decayed
exponentially with a 5~Myr timescale. However, with spectroscopic data
for only 2 such regions, it is difficult to draw useful conclusions
about the nature of the circumnuclear starburst in M100. Since
multi-object, near-infrared spectrographs are still under development,
long-slit spectroscopic observations like those presented here are of
necessity somewhat selective, but we have been able to secure spectra
for more than one-third of the compact knots discovered in Paper~I.

It is possible to draw some conclusions regarding the circumnuclear
star formation history of spiral galaxies from imaging in broad-band
filters.  However, broad-band colours are extremely sensitive to the
amount of reddening assumed, as well as to any emission from hot dust.
Narrow-band imaging of emission lines, including the use of
Fabry-P\'{e}rot etalons (e.g., Reunanen et al. 2000; Kotilainen et al.
2000), provides a
potentially more sensitive age diagnostic. Models have demonstrated
that the strength of the \Bgamma\ emission line equivalent width
begins to decline $\sim3$~Myr after the onset of a burst, but the rate
of decline will depend on the decay rate of the starburst, the IMF,
the metallicity, etc.

Over the same period, the equivalent width of the CO (2--0) bands will
grow to a maximum, and then decline, leading to an ambiguity in age
which is also model-dependent. Recently, the very usefulness of the CO
band strengths as an age-indicator for star formation has come under
renewed scrutiny (Origlia et al. 1999; Origlia \& Oliva 2000),
principally because of still-incomplete handling of the AGB phase of
red supergiant evolution by the evolutionary models, as well as the
competing effects of stellar effective temperature, surface gravity,
metallicity, and microturbulent velocity. Nevertheless, for starburst
ages greater than 5 and up to 100~Myr, we aim to demonstrate that the
{\em combination\/} of \Bgamma\ and CO (2--0) equivalent width
measurements can place quite useful constraints on the available
models, and serve as a powerful {\em relative\/} age discriminant.

In this paper, we present new low-resolution $K$-band spectra for many
more of the circumnuclear star-forming knots in M100, and determine
ages and burst parameters for them. We begin by describing our
spectroscopic observations and measurements in Sections~\ref{s:obs}
and \ref{s:res}. In Section~\ref{s:disc}, we compare our observed line
equivalent widths with starburst models, and use the derived ages to
search for radial and azimuthal trends. Our conclusions are contained
in Section~\ref{s:conc}, and in an Appendix, we highlight some aspects
of the CO spectroscopic index in nearby stars that are still not generally
appreciated.

\section[]{Observations and data reduction}\label{s:obs}

\begin{figure*}
\vspace{21cm}
\includegraphics{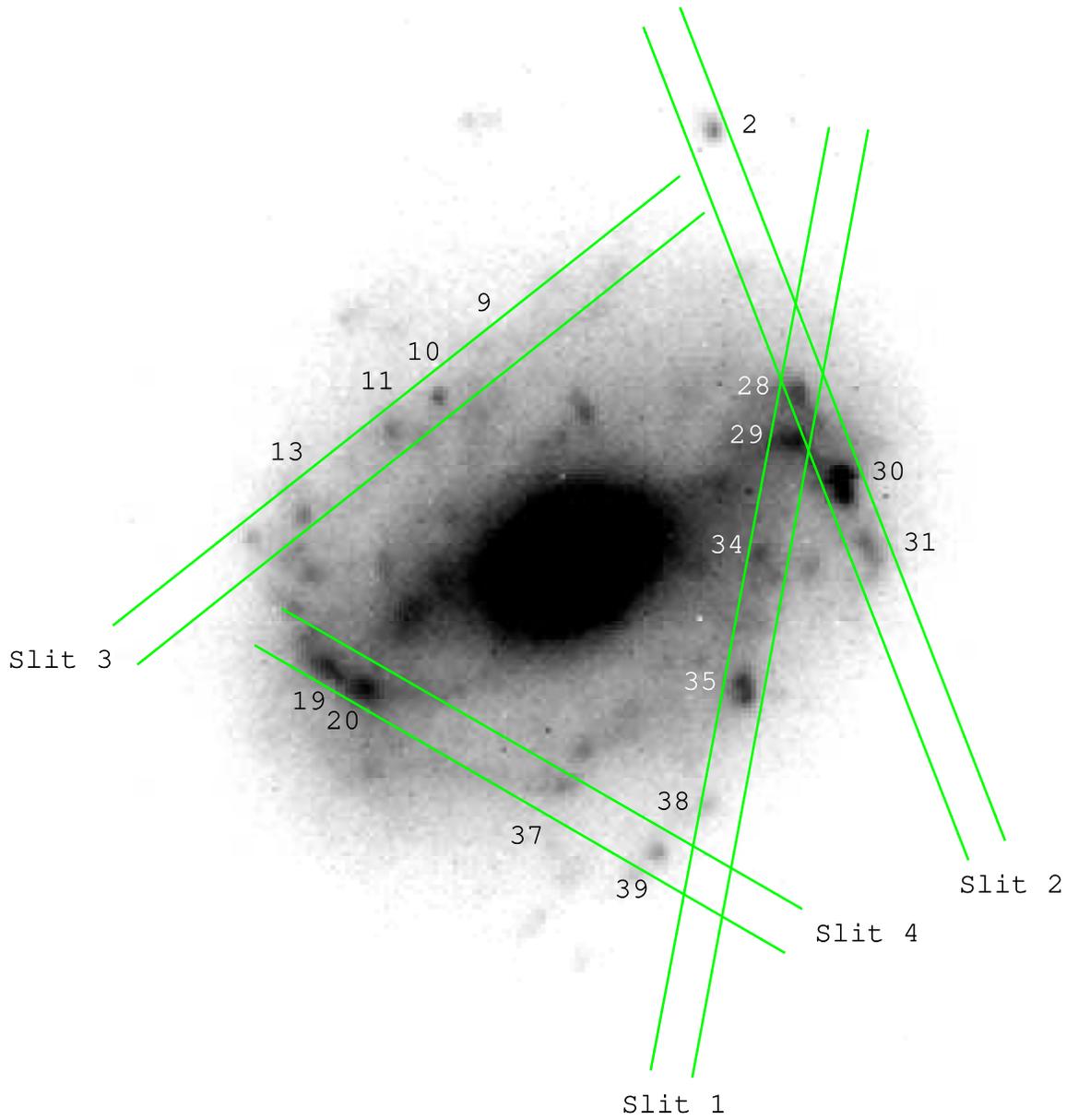}
\caption{Location of the CGS4 slit positions used, together with
identifications of the knots observed, marked on the $K$-band image
of the nuclear region of M100 from Ryder \& Knapen (1999). The 1.2~arcsec
slit width is shown to scale.}
\label{f:slits}
\end{figure*}

Longslit spectroscopy of the central region of M100 was carried
out over three nights from 1999~April~1-3 UT. Only limited data was
acquired the first night due to heavy cirrus; conditions were
photometric throughout the remaining two nights. The common-user near-IR
spectrograph CGS4 (Mountain et al. 1990) on UKIRT with its 0.61~arcsec pixel
scale, a 1.2~arcsec slit, and the 40~line~mm$^{-1}$ grating in first
order was employed, which delivers complete spectral coverage from
1.85--2.45~\mic\ at a resolving power of 450. Using the images
presented in Paper~I, a total of 4~slit position angles and nuclear
offsets were selected, covering at least 3~knots each; these are
marked on Figure~\ref{f:slits}.

After taking the observations presented in Paper~I, it became apparent
that the infrared-emitting central region of M100 fills no more than
one-third of the usable area of the detector array, and therefore,
nodding to blank sky was not really necessary. For the observations
presented here, sky subtraction was performed by sliding the objects
of interest a total of 60~rows (36.6~arcsec) along the slit in between
`object' and `sky' exposures. After setting the slit to the desired
position angle, a manual search for the $K$-band nuclear peak was
conducted, following which accurate offsets were applied by use of the
UKIRT crosshead to place the slit right at the locations marked on
Fig.~\ref{f:slits}. All observations of M100 were bracketed by similar
sequences on the nearby A4~V star BS~4632, to enable removal of some
of the telluric absorption features.

Data reduction was accomplished using a combination of tasks within
the Starlink {\sc cgs4dr} and NOAO {\sc iraf}\footnote{IRAF is
distributed by the National Optical Astronomy Observatories, which are
operated by the Association of Universities for Research in Astronomy,
Inc., under cooperative agreement with the National Science
Foundation.} packages. After flatfielding, the difference of each
object--sky pair was co-added into groups, with each grouped image
representing $\sim3000$~sec of on-source integration in the case of
the M100 spectra, or $\sim300$~sec for BS~4632. Any residual sky
absorption/emission was removed from this image by fitting a low-order
polynomial to galaxy-free regions within each row of the spatial
axis. The high signal-to-noise of the stellar continuum in BS~4632 was
used to define the width and curvature of suitable extraction
apertures for both the `positive' and `negative' beams. After
extracting matching apertures from an argon arc image, the spectra
from both beams were wavelength calibrated separately to an accuracy
of 0.2~nm. The spectrum of the negative beam was inverted, and then
co-added in wavelength space to that of the positive beam. The strong
\Bgamma\ absorption intrinsic to BS~4632 was interpolated over, but
this could not be done for \Palpha\ or for Br$\delta$ due to their
location near the trough of an H$_{2}$O absorption band.

Processing of the M100 knot spectra proceeded in a similar fashion,
except that multiple apertures needed to be defined for each slit
position. This was done in consultation with Fig.~\ref{f:slits}, but
the exact aperture centres and widths were defined independently for
each group. After extraction, wavelength calibration, and co-addition,
the spectrum of each knot was divided by the BS~4632 spectrum which
best matched the airmass of observation, and then multiplied by a
blackbody curve corresponding to the temperature (8380~K) and flux
density ($K=6.13$) of BS~4632. Table~\ref{t:slitpars} summarises the
observational parameters for each slit setting, including all of the
knots that each slit was expected to intersect to some degree. The
numbering sequence for the knots is as given in Table~1 of Paper~I.

In order to define a consistent continuum right out into the region of
extensive CO absorption beyond 2.3~\mic, we fitted a power-law of the
form $F_\lambda \propto \lambda^{\beta}$ to featureless sections of
the spectrum near 2.11 and 2.27~\mic\ (rest wavelength). After
normalising each spectrum by this fit, the equivalent widths of the
redshifted \Palpha, \Bgamma, and \H2\ lines were measured using the
SPLOT task in {\sc iraf}, together with two determinations of the
spectroscopic index \COsp.  The original definition of a CO spectroscopic
index by Doyon, Joseph \& Wright (1994a; hereafter DJW) has the
form

\vspace{5mm}
\begin{equation}
{\rm CO}_{\rm sp}^{\rm DJW} = -2.5\log \langle R_{2.36} \rangle
\label{eq:codjw}
\end{equation}
\vspace{5mm}

\noindent
where $R_{2.36}$ is the mean normalised intensity between 2.31 and
2.40~\mic\ in the rest frame of the galaxy (in the case of M100, this
wavelength interval corresponds to $2.322-2.412$~\mic).
Although the output of both the DJW and
the Leitherer et al. (1999; hereafter SB99) models use this definition
of \COsp, PDW advocated the adoption of a narrower wavelength interval
over which to integrate, namely $2.2931-2.3200$~\mic, which covers
just the region from the band head to near the end of the 2--0
band. After a comprehensive analysis, this range was found to offer
the best sensitivity to changes in luminosity class, and has also been
endorsed by Hill et al. (1999). From equation~7 of PDW, we get the
following relation between their definition of \COsp\ and the
equivalent width of CO (nm), measured over the rest wavelength
interval $2.2931-2.3200$~\mic:

\vspace{5mm}
\begin{equation}
{\rm CO}_{\rm sp}^{\rm PDW} = -2.5\log\left( \frac{16.8 -
{\rm EW(CO,nm)}}{\eta}\right)
\label{eq:pdw7}
\end{equation}
\vspace{5mm}

\noindent
At the redshift of M100 (1571~\kms; Knapen et al. 1993), this
wavelength interval becomes $2.305-2.332$~\mic. Note that PDW give
$\eta=16.6$, but our analysis in the Appendix shows that $\eta=16.9$
is more appropriate at our resolution.

\section[]{Results}\label{s:res}

Figure~\ref{f:spec} shows the spectra of the combined data for
three of the knots, on which the main spectral features that have
been measured are identified.
Even at this resolution, the redshift of M100 is just enough to
separate the \Palpha\ emission line from the anomalous bump at
1.876~\mic, which is a residual of the \Palpha\ absorption
in BS~4632. The (almost flat) continuum slope found in most
of the knot spectra helps greatly in minimising the error
that comes from fitting a power-law to the continuum and
extrapolating it across the CO band.

Table~\ref{t:knoteqw} is a compilation of the measured equivalent
widths and continuum power-law slopes $\beta$ for each of the knots
which could be clearly distinguished in the spectra. Because of the
tendency for CO$_{\rm sp}^{\rm PDW}$ to underestimate CO$_{\rm
sp}^{\rm DJW}$ as noted in the Appendix, the former has been
calculated using $\eta=16.9$ in equation~\ref{eq:pdw7}. In order to
gain a realistic estimate of the uncertainties, the equivalent width
measurements were performed on each of the 3 or 4 individually-reduced
groups (with exposure times of 2400--3800~sec), as well as the
co-added result. The mean equivalent width adopted is that measured
from the co-added spectra, while the error bars are based on the
dispersion in the individual measurements. As expected, this shows
that the uncertainties are dominated as much by the noise statistics and
the subjective choice of extraction aperture as by the measurement
errors and continuum fitting. Note that we have two independent
observations and measurements of knot~28 (with slit 1 and, less
optimally, slit 2), and except for the equivalent width of \H2, the
results are consistent within the error bars.

It is also apparent that the error bars on measurements over the
smaller PDW wavelength interval (their ``extended'' region) are
significantly larger than the error bars on measurements over the full
DJW range. These larger error bars, coupled with the need to then
apply an empirical transformation, leads us to favour the use of
CO$_{\rm sp}^{\rm DJW}$ for comparison with the models. While the PDW
definition undoubtedly is more robust than even narrower wavelength
regions, and is more sensitive to changes in the stellar population,
there appears to be little advantage to using it at our resolution.
Our conclusions in the next section regarding the relative ages of the
star-forming knots in M100 are unaffected by the choice of CO spectroscopic
index.

Since the knots all have CO$_{\rm sp}^{\rm DJW}$ in the range
0.20--0.28, we infer from Table~\ref{t:costars} that the stellar
populations of these knots are currently dominated by giants and/or
supergiants of type K5 or later. We do not observe values of
CO$_{\rm sp}^{\rm DJW}$ quite as high as was measured for the
`hotspots' K1 and K2 in Paper~I, which we attribute to the fact
that the original spectroscopy did not target any particular knot,
and that the continuum fitting was less rigorous.

% Reddening:  According to H99, to get a power law slope of -1.5 from a
% pure stellar (actually blackbody) law of -4 would require A_K~1.2
% mag, or A_V~11 mag (cf. A_V~5 mag in NGC 2903 from reddening the
% standards). Also, use task ``findred.cl'' to evaluate reddening
% and dominant spectral type cf. a standard star

\section[]{Discussion}\label{s:disc}

\subsection[]{The Ages of the Starburst Knots}\label{s:ages}

Figure~\ref{f:cobg1} presents our results as a plot of CO
spectroscopic index {\em vs.} the \Bgamma\ equivalent width for the 16
knots (the two independent observations of knot 28 have been
averaged), together with the loci of evolutionary synthesis models for
starbursts as calculated by DJW, and used by PDW in age-dating the
nuclear starburst in M83. There appears to be a well-defined sequence
in the line strengths among the bulk of the knots, which parallels
quite closely the starburst model with the shortest burst timescale
$\tau$ (i.e. the ``quasi-instantaneous'' burst). Reading off the
fiducial marks on this model indicates ages for these knots of between
8 and 10~Myr.

There is of course a selection effect here: the \Bgamma\ equivalent
width must be at least 0.2~nm to be detectable, so we can only put a
lower age limit on knot~13. Similarly, there could be circumnuclear
star-forming knots younger than 5~Myr around M100 which would be
missed by our NIR imaging and spectroscopy, since there has not been
sufficient time for the most massive stars to evolve to red
supergiants. Nevertheless the observed upper envelope of \Bgamma\
equivalent widths just below 1~nm suggests that there was an extended
pause in star formation events $\sim8$~Myr (on this model)
ago. Furthermore, any bursts decaying at a rate slower than 5~Myr, and
aged between 10 and 30~Myr ought also to be detectable, but none are
seen.

We have also plotted our results on the closest equivalent SB99 models
(Figure~\ref{f:cobg2}). The agreement between observations and the
models is not as good as for the DJW models, but the two evolutionary
tracks shown are the only two that come anywhere close to achieving
the required values of CO$_{\rm sp}^{\rm DJW}$.  The first model is
for an instantaneous burst of star formation (ISF), in which the IMF
has a Salpeter-like form (slope $\alpha=2.35$ and upper mass cutoff of
100~M$_{\odot}$) and a metallicity of $2\times$ solar ($Z=0.04$). The
alternative model is one which has a constant star formation rate
(CSF) of 1~M$_{\odot}$~yr$^{-1}$, a steeper IMF ($\alpha=3.30$ and
upper mass cutoff of 100~M$_{\odot}$) and a metallicity of
twice-solar.

A high central metallicity in M100 is supported by \HII~region
spectroscopic abundance measurements by Skillman et al. (1996), who
found $12+\log({\rm O}/{\rm H})\sim9.3$ near the nucleus, or over
$2.5\times$~solar metallicity. The tendency of existing models to
predict CO spectroscopic indices which are too weak is a known problem
(e.g., Devost \& Origlia 1998; Origlia et al. 1999), and appears to be
related to uncertainties in the effective temperatures and
microturbulent velocities of the red supergiants, as well as the
fraction of time these stars spend as blue supergiants during their
core helium burning phase.  Nevertheless, the evolutionary trend of
the ISF model parallels quite closely the distribution of the knots,
and even the age range of $6.5-7.0$~Myr is not inconsistent with that
indicated by the DJW models, considering that the ISF model has $\tau
< 0.1$~Myr.

Although the (quasi-)instantaneous burst models fit the observed trend
of knot line indices quite well, both Figures~\ref{f:cobg1} and
\ref{f:cobg2} leave open the possibility that a few of the knots could
equally well just be due to steady (or only slowly declining)
continuous star formation for the past 50--100~Myr. The same CSF model
plotted in Fig.~\ref{f:cobg2} predicts that in this age range, the
knots should have colours $(J-H)=0.48\pm0.02$ and $(H-K)=0.24\pm0.02$,
while the ISF model in the same figure predicts $(J-H)=0.46\pm0.05$
and $(H-K)=0.22\pm0.04$ for the age range 6.6--7.0~Myr. Each of these
cases is fully consistent with the median of the dereddened knot
colours, as plotted in Fig.~2 of Paper~I, emphasising how ineffective
broadband near-infrared colours are in distinguishing bursts from
continuous star formation. As we show in Section~\ref{s:molh2}, even
knowledge of the H$_{2}$ 1--0 S(1) to \Bgamma\ ratio still does not
allow for an unambiguous distinction between instantaneous burst and
continuous star formation models.

\subsection[]{Other Hydrogen Lines}\label{s:h2pa}

\subsubsection[]{Pa$\alpha$}\label{s:palpha}

\begin{figure*}
\vspace{21cm}
\includegraphics{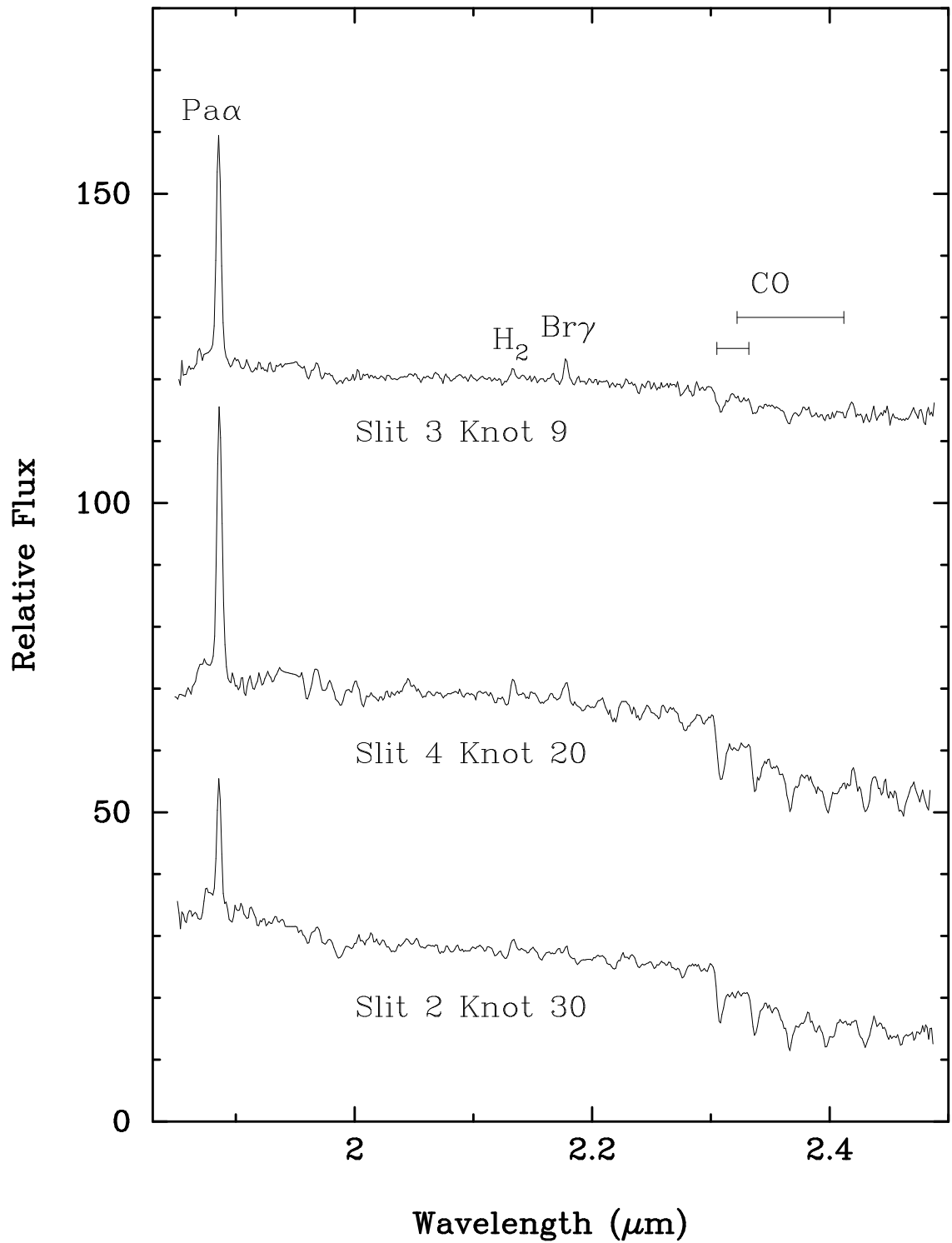}
\caption{Reduced spectra for three of the knots in M100, illustrating the
relative changes in line strengths going from a relatively young knot (Knot 9,
age $\sim8.4$~Myr) to one of the oldest knots (Knot 30, age $\sim9.6$~Myr),
via one of intermediate age (Knot 20, age $\sim8.8$~Myr). The main spectral
features are marked,
as well as the two wavelength regions over which the CO equivalent width
was measured: ({\em upper line}) the range defined by DJW, and ({\em lower
line}) the narrower range adopted by PDW. The bump on the blue wing of
the \Palpha\ lines is a residual of the \Palpha\ 1.876~\mic\ absorption
intrinsic to the telluric standard star. The residual due to Br$\delta$
at 1.945~\mic\ has been patched over in these spectra.}
\label{f:spec}
\end{figure*}

As Fig.~\ref{f:spec} shows, the \Palpha\ line at the redshift of M100
(1571~\kms) is clearly resolved from the rest-frame \Palpha\
absorption in BS~4632. Thanks to the low water vapour column over
Mauna Kea, measurements of \Palpha\ are quite feasible despite the low
atmospheric transmission at this wavelength, although the continuum
level in this region is not as well fit by the power law. For this
reason, the equivalent widths in Table~\ref{t:knoteqw} have been
measured with respect to the base of the line profile, rather than the
continuum as defined by an extrapolation of the power law with index
$\beta$.

In Figure~\ref{f:bgpa} we have plotted the equivalent width of the
\Palpha\ line against that of the \Bgamma\ line. For Case~B
recombination of hydrogen at $T=10^{4}$~K (e.g. Table~4.2 of
Osterbrock 1989), we would expect a ratio $I({\rm P}\alpha)/ I({\rm
Br}\gamma)=12.52$, as indicated by the dashed line (more metal-rich
gas will be cooler, but even at $T=5\times10^{3}$~K, the ratio is only
slightly higher at 12.86). The fact that the \Palpha\ line is consistently
weaker than \Bgamma\ would predict is usually attributed to reddening.
However, the mean amount of extinction required to explain this
difference (adopting the interstellar extinction curve from Cardelli,
Clayton \& Mathis 1989) is $A_{V}\sim 16$~mag, or $A_{K}\sim1.8$~mag.
This is a factor of 4 larger than the maximum extinction that was
derived from the near-IR colour excesses of the knots in Paper~I.

The equivalent width of an emission line ought to be independent of
reddening, but this assumes that both the continuum flux and the line
emission at a given wavelength originate from the same physical
region; if the red supergiants had cleared much of the dust that
shrouded their birth, then the stellar and nebular fluxes could
experience quite different amounts of reddening. Alternatively, our
corrections for \Palpha\ absorption in the standard star, and for
H$_{2}$O absorption in our atmosphere, may be inadequate. An added
complication is the presence of extended, diffuse \Palpha\ emission in
the nuclei of many early-type galaxies (B\"{o}ker et al. 1999;
Alonso-Herrero et al. 2001). Clearly, despite being an order of
magnitude stronger than \Bgamma, there are still a number of
difficulties with using ground-based observations of the \Palpha\ line
as a star formation and reddening diagnostic.

\subsubsection[]{H$\alpha$}\label{s:halpha}

The \Halpha\ narrow-band imaging and Fabry-P\'{e}rot kinematics
presented by Knapen et al. (1995a; 2000) reveal four main complexes of
\HII~regions forming an almost complete nuclear ring, coincident with
that outlined by the $K$-band knots. Two of these complexes are
associated with the major `hot-spots' K1 and K2 at either end of the
bar. On the basis of reduced dust obscuration, they suggested that K1
ought to be slightly older than K2, and indeed, our age analysis
confirms that the aggregate of knots which make up K1 (knots 28-31) is
slightly older than that of K2 (knots 19 and 20). Complexes H$\alpha3$
and H$\alpha4$, along the bar minor axis, are brighter than K1 and K2
in \Halpha, but dimmer in $K$, consistent with their being less evolved.
As discussed in Section~\ref{s:ages}, the youngest knots are found near
H$\alpha3$ and H$\alpha4$, providing further support for the qualitative
age sequence of Knapen et al.

\subsubsection[]{Molecular hydrogen}\label{s:molh2}

Molecular hydrogen emission in the H$_{2}$ 1--0 S(1) line at a rest
wavelength of 2.122~\mic\ is commonly observed in star-forming
environments, but the nature of its excitation (shocks {\em vs.}  UV
fluorescence) is still somewhat ambiguous (e.g., Doyon, Wright \&
Joseph 1994b; Puxley, Ramsay~Howat \& Mountain 2000). When we plot the
H$_{2}$ 1--0 S(1) line strengths against \Bgamma\
(Figure~\ref{f:bgh2}), the two are clearly correlated (just as was
demonstrated for the nuclei of a large sample of star-forming galaxies
by Puxley, Hawarden \& Mountain 1990), suggesting a close link between
molecular hydrogen line emission and star formation in these
knots. Table~\ref{t:knoteqw} lists the ratios of the H$_{2}$ 1--0 S(1)
to \Bgamma\ equivalent widths, and these are predominantly in the
range $0.4-0.9$, very similar to the Puxley et al. (1990)
ratios.
% Their geometrical modeling suggested that line ratios in this
%range were best accounted for by an ensemble of photodissociation
%regions (PDRs) surrounding individual hot stars and their \HII~regions
%(as opposed to either a cluster embedded in a single PDR, or to
%molecular clouds bathed in an ambient UV field).

Using the same modeling framework as used for the evolutionary tracks
in Fig.~\ref{f:cobg1}, Doyon et al. (1994b) showed that the
S(1)/\Bgamma\ ratio would not exceed 0.4 until at least 60~Myr after
the onset of the burst, for $\tau=20$~Myr. The short ($\tau=1$~Myr)
bursts favoured by Fig.~\ref{f:cobg1} might be expected to achieve the
observed S(1)/\Bgamma\ ratios in Table~\ref{t:knoteqw} even earlier
than 60~Myr. Their model assumes however that outflows from young
stellar objects (YSOs) and from supernova remnants (SNRs) produce all
of the H$_{2}$ line flux, whereas both UV fluorescence of
photodissociation regions by massive stars, and large-scale shocks
from the action of the spiral density wave, may also be significant
sources in the centre of M100.

For comparison, their model with continuous star formation (like the
CSF model of SB99 in Fig.~\ref{f:cobg2}) would barely exceed a ratio
of 0.15, even after 100~Myr of evolution.  Vanzi \& Rieke (1997)
measured ratios of this order in a sample of blue dwarf galaxies
which, while representing perhaps the purest form of a starburst, may
also be subject to the unknown effects of low metallicity.  Thus, even
a combined knowledge of the knot near-IR colours, their \Bgamma,
H$_{2}$ and CO equivalent widths is still not sufficient to
definitively exclude the possibility that at least some of the
circumnuclear knots in M100 have been continuously forming stars for
as long as 100~Myr.

\subsection[]{Age trends}\label{s:trends}

Table~\ref{t:ages} lists the ages derived for each of the knots which
lie on, or very near to the $\tau=1$~Myr locus of DJW in
Fig~\ref{f:cobg1}. Four of the knots (2, 13, 28, and 38) lie well off
this sequence, and it would appear that the bursts powering them are
declining at a different rate from all the rest. Clearly, ages derived
from the SB99 ISF model in Fig.~\ref{f:cobg2} would be slightly younger,
but the relative age sequence will still be the same. The location of
each knot within the deprojected disk of M100 has been calculated
using the disk orientation parameters ($i=27^{\circ}$,
$\phi=153^{\circ}$) derived by Knapen et al. (1993) from their \HI\
kinematic analysis.  In Figure~\ref{f:ageplots}, we plot these ages
against both radial distance and position angle in the plane of
M100. In this sense, position angle increases counter-clockwise, with
the origin lying midway between knots 20 and 37 in Fig~\ref{f:slits}.

We consider first the azimuthal age distribution. The knots at either
end of the bar (19, 20 and 29) are all about the same age,
$\sim8.8$~Myr. The oldest knots (30, 31, 34, and 35) are located just
south of the western end of the inner bar, while the youngest knots
(9, 37 and 39) lie close to the bar minor axis, leading (or lagging)
the oldest knots by $90^{\circ}$.  Indeed, beginning at position angle
$70^{\circ}$, there is the suggestion of an age sequence spanning at
least a Myr as one circles the nuclear `ring'. A simple linear least
squares fit (Press et al. 1992) to the age distribution, starting at
position angle $70^{\circ}$, yields a Pearson's $r$ value of $-0.75$
and Student's $t$ probability of 0.005, the small value of which
implies a significant correlation between position angle around the
ring, and burst age. However, one could just as easily point to the
appearance of {\em two\/} age minima (at position angles $50^{\circ}$
and $230^{\circ}$), but our limited sampling prevents us from
identifying if there is a corresponding second age maximum near
$300^{\circ}$. Nevertheless, Fig.~\ref{f:ageplots} provides the first
evidence that rather than being a stochastic process, {\em
circumnuclear star formation may be sequentially triggered}.

At the radial distance of these knots, the orbital period given by the
rotation curve of Knapen et al. (1993) is $\sim20$~Myr, an order of
magnitude larger than the observed age spread. Similarly, the bar
pattern speed (determined independently by Knapen et al. (1995b) and
by Wada et al. (1998)), $\Omega_{\rm p} \sim 65$~\kms~kpc$^{-1}$ leads
to an even longer timescale. The maximum streaming velocities observed
in \Halpha\ and CO by Knapen et al. (2000) around the circumnuclear
region are of order 40~\kms, too slow to causally connect the oldest
and youngest star-forming knots. Thus, it cannot be the rotation of the
bar, or bulk motion of the gas, which directly triggers the star
formation.

The detailed nuclear morphology of M100, as pieced together from
optical, near-infrared, \Halpha, and CO imaging, is analysed in great
detail by Knapen et al. (1995a, b; 2000). They constructed a numerical
model including gas, stars, and star formation, which successfully
replicated many of the main features, including (1) the 1~arcmin
stellar bar; (2) a pair of offset curved shocks, and streaming motions
along the nuclear `ring', both signatures of a global density wave
induced by this bar; and (3) four distinct compression zones, two of
which are close to the bar minor axis (where local maxima in the CO
and \Halpha\ emission are observed), and two near the ends of the bar
where the spiral armlets are observed to switch from leading to
trailing. As shown in fig.~15 of Knapen et al. (1995b), star formation
at the ends of the bar in their model is seen to precede that along
the minor axis (just as Fig.~\ref{f:ageplots} suggests), but again
the timescales involved are rather longer than the observed age spread.

One scenario worth considering would be an outward propagating
shock-wave, which reaches the ends of the bar first, then sets off the
knots along the bar minor axis a short while later, resulting in a
radial age gradient. As Fig.~\ref{f:ageplots} shows, the oldest knots
do tend to be at larger radii, but then so too is the youngest, so the
evidence is less compelling (for a linear least squares fit, $r=0.22$
and $t=0.5$). What makes this model attractive though, is that a shock
wave velocity of only $\sim200$~\kms\ is required to explain the
observed age spread, compared with up to 1000~\kms\ for a density
wave-driven shock propagating around the ring. In addition, a one-off
outburst such as this makes it easier to account for the apparent
recent pause in star formation events, whereas a persistent phenomenon
(such as a spiral density wave for instance) would be expected to have
produced a much larger spread in burst ages than is observed in M100.

\section{Conclusions}\label{s:conc}

New longslit $K$-band spectroscopy of the circumnuclear region of M100
has been obtained, targeting specifically the compact `knots'
identified from high-resolution near-IR imaging in Paper~I. We have
shown that even at comparatively low resolution ($R\sim450$), it is
possible to discern age differences of as little as 0.2~Myr in
circumnuclear star forming regions by comparing the equivalent widths
of the \Bgamma\ emission line with the CO 2.29~\mic\ absorption
band. While the primary age sequence diagnostic is the CO
spectroscopic index, complementary measurements of the \Bgamma\ and
H$_{2}$ 1--0 S(1) lines allows us to distinguish short-duration,
rapidly declining bursts from continuous star formation. Comparison
with starburst models from the literature indicates that the bulk of
the circumnuclear star formation events in M100 are best accounted for
by the former, with decay timescales of $\sim1$~Myr.

We find the strongest evidence yet that circumnuclear star formation
in M100 is sequentially triggered in an azimuthal or radial sense (or
perhaps both). While the absolute ages of the knots are highly
model-dependent, the age spread is unambiguous. The youngest and
oldest knots are quite spatially distinct, but the physical
separations and timescales involved are such that only shocks with
velocities of a few hundreds of kilometres per second can causally
connect them. The presence of curved dust lanes bisecting the
circumnuclear ring verifies the existence of large-scale shocks
in this region, most likely driven by a bar-induced spiral density
wave. However, the symmetry of the age distribution is such that
a simple, radially-propagating shock may just as easily explain the
situation, though no source can yet be identified.

To date, we have secured spectra for 16 of the 43 knots identified in
Paper~I, and of these, only 12 are suitable for constraining the age
distribution. To confirm our suspicion of an azimuthal age gradient
(or even a cycle) in M100 will require better sampling of some of the
fainter knots in M100 (something for which the new generation of
near-infrared multi-object spectrographs and integral-field units will
be ideally suited). Few galaxies have had their circumnuclear
kinematics and star formation studied in as much detail as M100 has,
but more are clearly needed if we are to understand the role of the
bar in influencing both.

\section*{Acknowledgments}

\begin{figure*}
\vspace{21cm}
\includegraphics{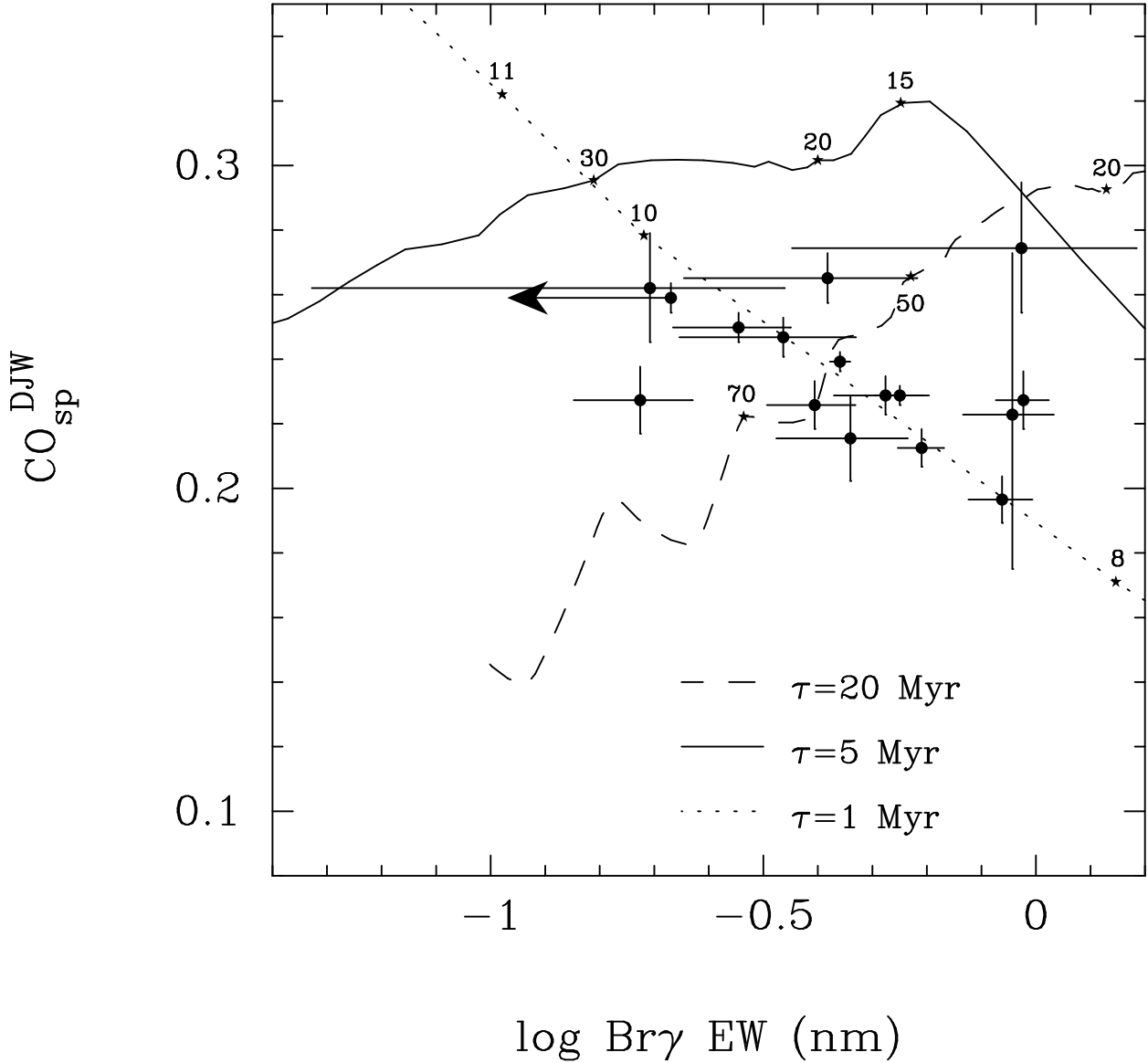}
\caption{Observed CO spectroscopic indices plotted against the \Bgamma\
equivalent widths for 16 of the star-forming knots in M100. The two
independent observations of knot~28 in Table~\protect{\ref{t:knoteqw}}
have been combined with a weighted average, to give the point at
$(-0.73, 0.23)$. Also shown in this plot are the evolutionary tracks
for three starburst models as presented
in Fig.~3 of PDW, with exponentially-decreasing star formation timescales
$\tau=1$, 5, and $20\times10^{6}$~yr. The points marked on the tracks
indicate the time taken in Myr to reach that point from the onset of
the burst.}
\label{f:cobg1}
\end{figure*}

\begin{figure*}
\vspace{21cm}
\includegraphics{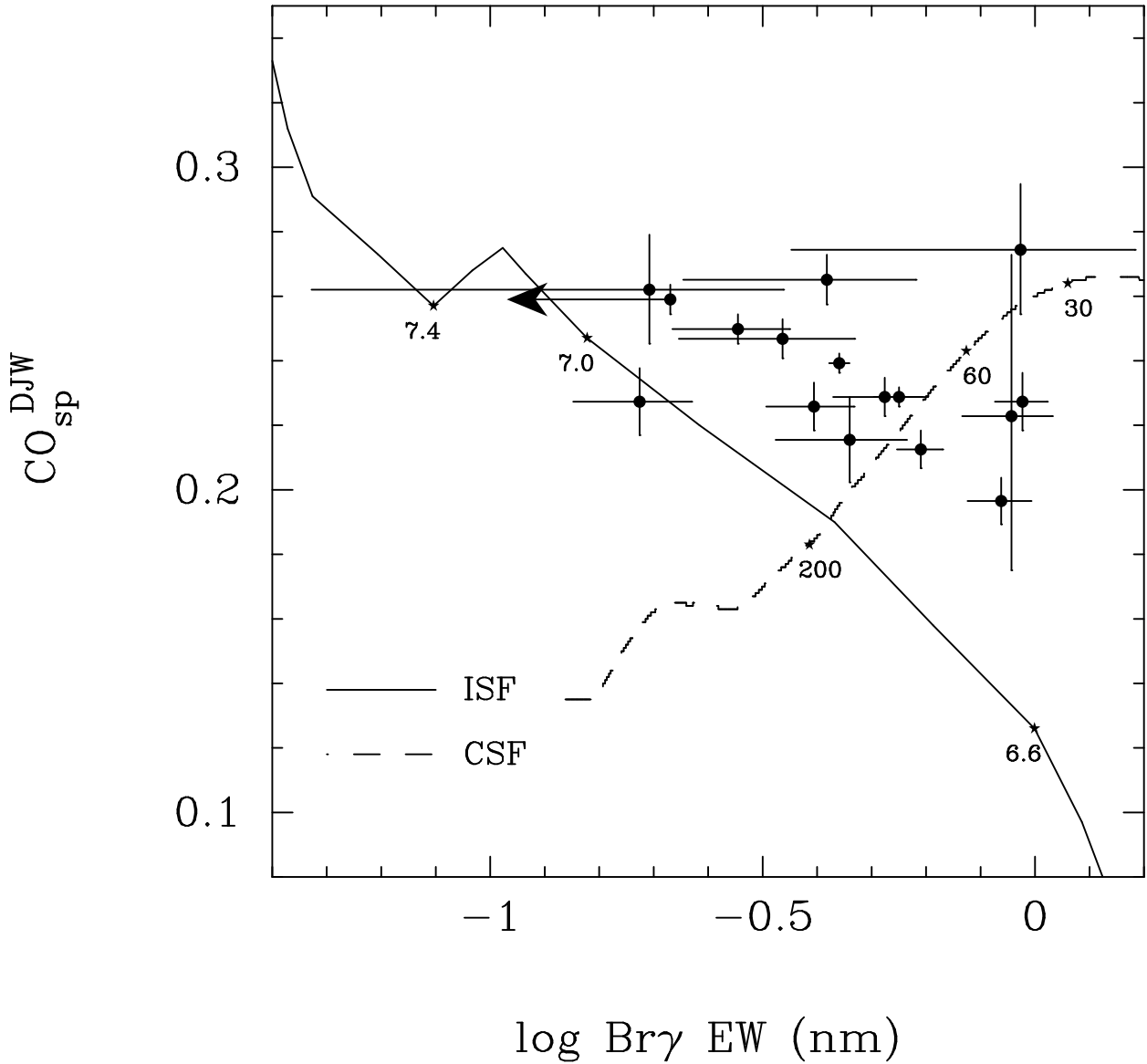}
\caption{As for Figure~\protect{\ref{f:cobg1}}, but this time using two
evolutionary tracks from Leitherer et al. (1999) models. The track marked
``ISF'' is for a $Z=0.04$, $\alpha=2.35$, M$_{\rm up}=100$~M$_
{\odot}$ instantaneous burst model, while the one marked ``CSF'' is
for a $Z=0.04$, $\alpha=3.30$, M$_{\rm up}=100$~M$_{\odot}$
model with continuous star formation at a rate of 1~M$_{\odot}$~yr$^{-1}$.
The points marked on the tracks
indicate the time taken in Myr to reach that point from the onset of star
formation.}
\label{f:cobg2}
\end{figure*}

\begin{figure*}
\vspace{21cm}
\includegraphics{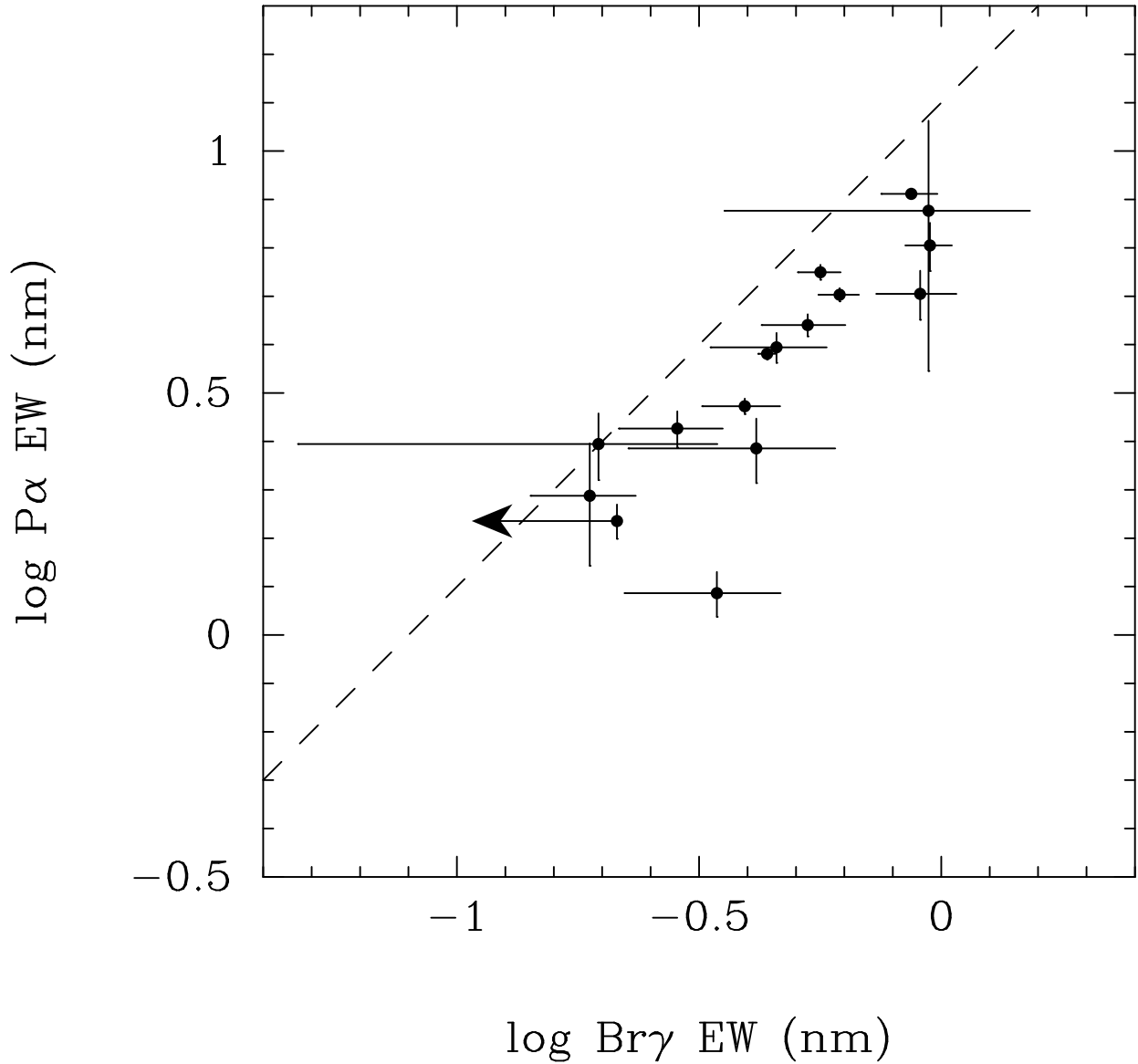}
\caption{The equivalent width of the \Palpha\ emission line from each
of the knots plotted against the \Bgamma\ equivalent width. The relationship
expected from Case~B recombination line theory is shown by the dashed line.
The two independent observations of knot~28 in Table~\protect{\ref{t:knoteqw}}
have been combined with a weighted average, to give the point at
$(-0.73, 0.29)$.}
\label{f:bgpa}
\end{figure*}

\begin{figure*}
\vspace{21cm}
\includegraphics{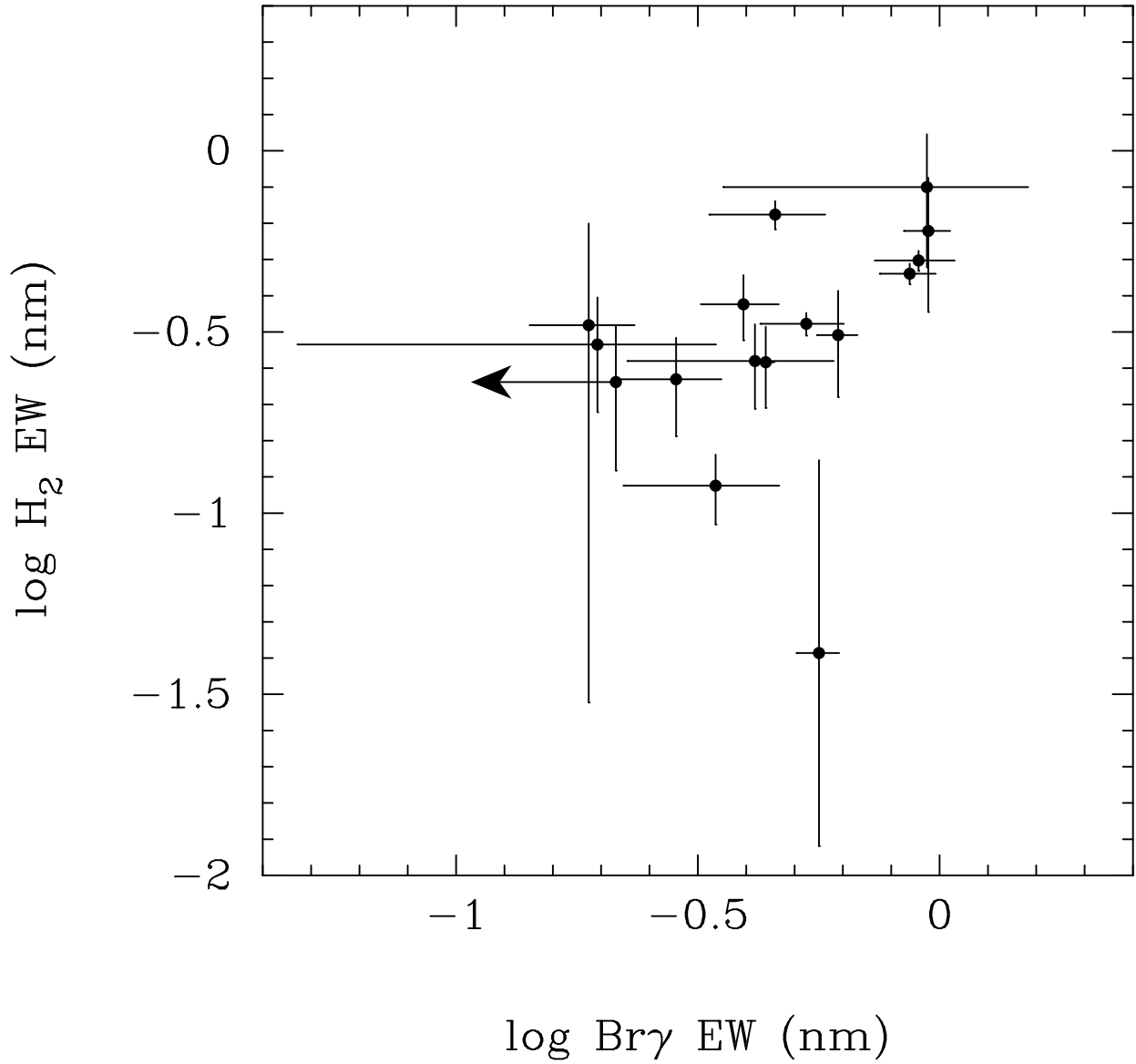}
\caption{The equivalent width of the H$_{2}$ 1--0 S(1) emission line from each
of the knots plotted against the \Bgamma\ equivalent width. The two independent
observations of knot~28 in Table~\protect{\ref{t:knoteqw}} have been combined
with a weighted average, to give the point at $(-0.73, -0.48)$.}
\label{f:bgh2}
\end{figure*}

\begin{figure*}
\vspace{21cm}
\includegraphics{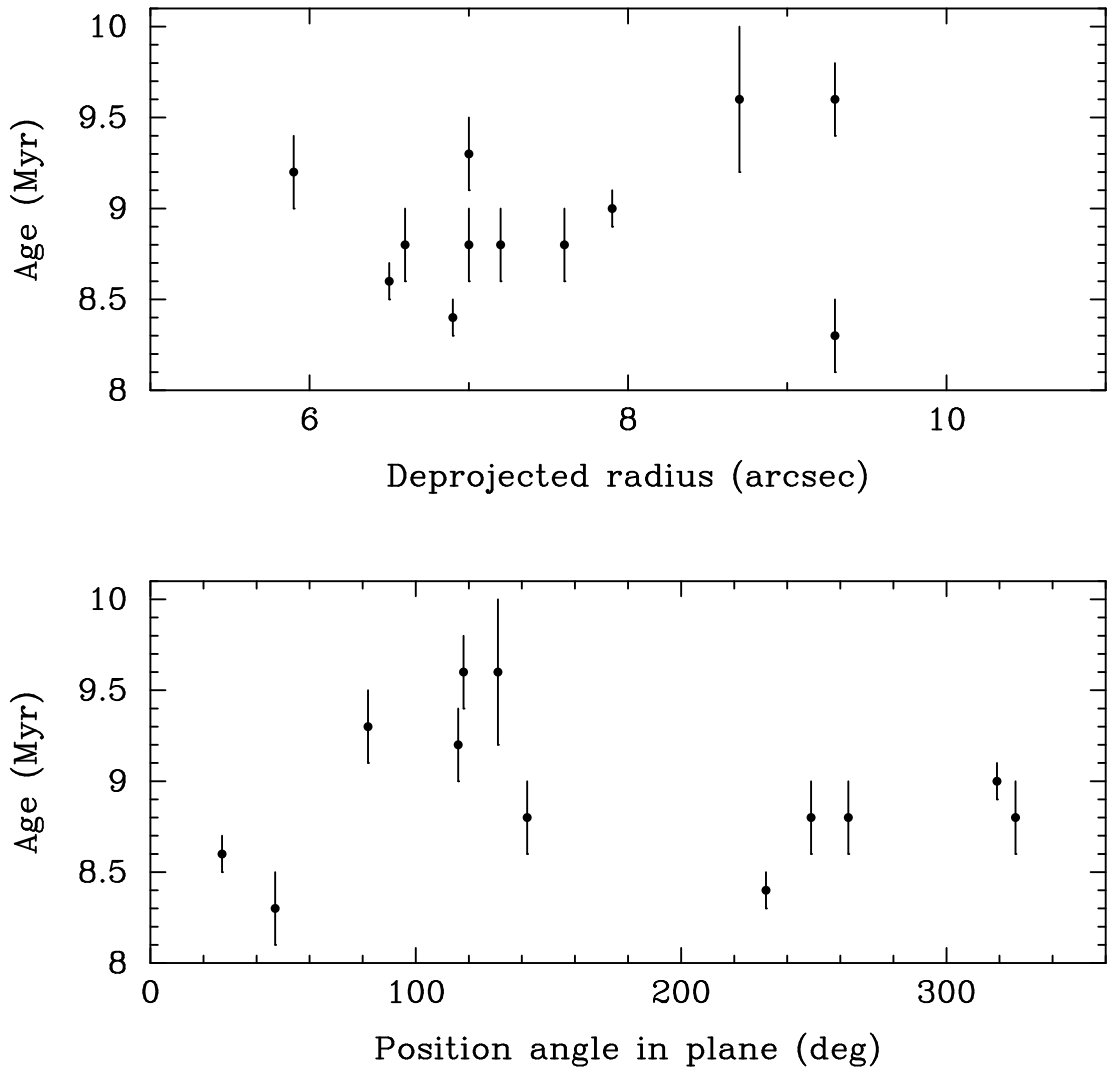}
\caption{The derived ages for the 12 knots in M100 which appear to
best follow the evolutionary trend of the $\tau=1$~Myr burst locus in
Fig.~\protect{\ref{f:cobg1}}, plotted against deprojected radius ({\em top})
and azimuth ({\em bottom}) in the disk of M100. The zero-point of the
position angle scale corresponds to the kinematical major axis
$\phi=153^{\circ}$ (roughly halfway between knots 20 and 37 in
Fig.~\protect{\ref{f:slits}}), and position angle increases going
counter-clockwise.}
\label{f:ageplots}
\end{figure*}

This research has made use of the NASA/IPAC Extragalactic Database
(NED), which is operated by the Jet Propulsion Laboratory, California
Institute of Technology, under contract with the National Aeronautics
and Space Administration, and of NASA's Astrophysics Data System (ADS)
abstract service. The United Kingdom Infrared Telescope is operated by
the Joint Astronomy Centre on behalf of the U.K. Particle Physics and
Astronomy Research Council.  We are grateful to R.~Doyon and P.~Puxley
for their advice and for providing the starburst model results in
Fig.~\ref{f:cobg1}. The comments of an anonymous referee helped our
interpretation of some of these results.

\newpage

\begin{table}
 \caption{CGS4 Slit Parameters for M100}
 \label{t:slitpars}
 \begin{tabular}{crcr}
\hline
Slit & \multicolumn{1}{c}{P.A.} & {Knots} & \multicolumn{1}{c}{Total Exp.
(sec)} \\
\hline
1   &  169.6  &  28, 29, 34, 35, 38  &  10080  \\
2   &   17.9  &  2, 28, 30, 31       &  14880$^{a}$  \\
3   &  128.3  &  9, 10, 11, 13, 14   &  7200  \\
4   &   60.9  &  18. 19, 20, 37, 39  &  8640  \\
\hline
\end{tabular}
\medskip
%~~\\
\begin{flushleft}
$^{a}$Includes 5520~sec through thick cirrus.\\
\end{flushleft}
\end{table}

\newpage

\begin{table*}
 \caption{Line Equivalent Widths and CO Spectroscopic Indices for the
Star-forming Knots in M100}
 \label{t:knoteqw}
 \begin{tabular}{rccccccc}
\hline
Knot  & $\beta$      & --EW(P$\alpha$) & --EW(H$_{2}$ 1--0 S(1)) &
 --EW(Br$\gamma$)  & S(1)/Br$\gamma$   & CO$_{\rm sp}^{\rm DJW}$ &
CO$_{\rm sp}^{\rm PDW}$  \\
      &              &      (nm)       &      (nm)               & 
      (nm)         &                &                  &          \\
\hline
\multicolumn{8}{c}{Slit 1} \\
\hline
 28   & $-1.2\pm0.2$ &  $1.63\pm0.27$  &  $0.09\pm0.06$          &
 $0.20\pm0.06$     &    $0.45\pm0.44$  &  $0.23\pm0.01$  & $0.23\pm0.02$  \\  
 29   & $-1.0\pm0.1$ &  $2.97\pm0.11$  &  $0.38\pm0.08$          &
 $0.39\pm0.07$     &    $0.97\pm0.38$  &  $0.23\pm0.01$  & $0.22\pm0.01$  \\  
 34   & $-1.1\pm0.1$ &  $1.22\pm0.13$  &  $0.12\pm0.03$          &
 $0.34\pm0.12$     &    $0.35\pm0.21$  &  $0.25\pm0.01$  & $0.25\pm0.01$  \\  
 35   & $-1.1\pm0.1$ &  $2.67\pm0.23$  &  $0.23\pm0.07$          &
 $0.29\pm0.07$     &    $0.79\pm0.43$  &  $0.25\pm0.01$  & $0.25\pm0.01$  \\  
 38   & $-0.9\pm0.2$ &  $6.38\pm0.73$  &  $0.60\pm0.24$          &
 $0.95\pm0.10$     &    $0.63\pm0.32$  &  $0.23\pm0.01$  & $0.27\pm0.08$  \\  
\hline
\multicolumn{8}{c}{Slit 2} \\
\hline
  2   & $-0.4\pm0.2$ &  $7.53\pm4.02$  &  $0.79\pm0.32$          &
 $0.94\pm0.58$     &    $0.84\pm0.84$  &  $0.27\pm0.02$  & $0.32\pm0.05$  \\ 
 28   & $-1.1\pm0.3$ &  $2.06\pm0.45$  &  $0.42\pm0.21$          &
 $0.22\pm0.01$     &    $1.91\pm1.04$  &  $0.23\pm0.01$  & $0.22\pm0.01$  \\
 30   & $-1.1\pm0.2$ &  $2.48\pm0.39$  &  $0.29\pm0.10$          &
 $0.20\pm0.15$     &    $1.45\pm1.45$  &  $0.26\pm0.02$  & $0.26\pm0.02$  \\
 31   & $-1.1\pm0.2$ &  $2.43\pm0.37$  &  $0.26\pm0.07$          &
 $0.42\pm0.19$     &    $0.62\pm0.45$  &  $0.27\pm0.01$  & $0.27\pm0.03$  \\
\hline
\multicolumn{8}{c}{Slit 3} \\
\hline
  9   & $-1.0\pm0.1$ &  $8.16\pm0.09$  &  $0.46\pm0.03$          &
 $0.87\pm0.12$     &    $0.53\pm0.11$  &  $0.20\pm0.01$  & $0.18\pm0.01$  \\
 10   & $-0.8\pm0.2$ &  $5.62\pm0.20$  &  $<0.04$                &
 $0.56\pm0.06$     &    $<0.07$        &  $0.23\pm0.01$  & $0.25\pm0.03$  \\
 11   & $-0.8\pm0.1$ &  $3.93\pm0.28$  &  $0.67\pm0.06$          &
 $0.46\pm0.12$     &    $1.46\pm0.51$  &  $0.22\pm0.01$  & $0.22\pm0.02$  \\
 13   & $-0.9\pm0.1$ &  $1.72\pm0.14$  &  $<0.23$                &
 $<0.21$           &    $<1.1$         &  $0.26\pm0.01$  & $0.27\pm0.04$  \\
\hline
\multicolumn{8}{c}{Slit 4} \\
\hline
 19   & $-1.2\pm0.1$ &  $3.81\pm0.10$  &  $0.26\pm0.07$          &
 $0.44\pm0.02$     &    $0.59\pm0.19$  &  $0.24\pm0.01$  & $0.23\pm0.02$  \\
 20   & $-1.0\pm0.1$ &  $4.37\pm0.23$  &  $0.33\pm0.02$          &
 $0.53\pm0.10$     &    $0.62\pm0.15$  &  $0.23\pm0.01$  & $0.23\pm0.01$  \\
 37   & $-1.1\pm0.1$ &  $5.05\pm0.16$  &  $0.31\pm0.10$          &
 $0.62\pm0.06$     &    $0.50\pm0.21$  &  $0.21\pm0.01$  & $0.22\pm0.02$  \\
 39   & $-1.1\pm0.2$ &  $5.07\pm0.59$  &  $0.50\pm0.03$          &
 $0.91\pm0.17$     &    $0.55\pm0.14$  &  $0.22\pm0.01$  & $0.26\pm0.04$  \\
\hline
\end{tabular}
\end{table*}

\newpage

\begin{table}
 \caption{Knot Locations and $\tau=1$~Myr Burst Ages}
 \label{t:ages}
 \begin{tabular}{rrrc}
\hline
Knot & \multicolumn{1}{c}{R} & \multicolumn{1}{c}{$\theta$} &
Age \\
     & \multicolumn{1}{c}{(arcsec)} & \multicolumn{1}{c}{($^{\circ}$)} &
(Myr) \\
\hline
 2 & 13.0 & 190 & $\ldots$      \\
 9 &  6.9 & 232 & $8.4 \pm 0.1$ \\
10 &  6.6 & 249 & $8.8 \pm 0.2$ \\
11 &  7.0 & 263 & $8.8 \pm 0.2$ \\
13 &  8.6 & 286 & $\ldots$      \\
19 &  7.9 & 319 & $9.0 \pm 0.1$ \\
20 &  7.2 & 326 & $8.8 \pm 0.2$ \\
28 &  8.2 & 149 & $\ldots$      \\
29 &  7.6 & 142 & $8.8 \pm 0.2$ \\
30 &  8.7 & 131 & $9.6 \pm 0.4$ \\
31 &  9.3 & 118 & $9.6 \pm 0.2$ \\
34 &  5.9 & 116 & $9.2 \pm 0.2$ \\
35 &  7.0 &  82 & $9.3 \pm 0.2$ \\
37 &  6.5 &  27 & $8.6 \pm 0.1$ \\
38 &  8.6 &  59 & $\ldots$      \\
39 &  9.3 &  47 & $8.3 \pm 0.2$ \\
\hline
\end{tabular}
\end{table}

\clearpage

\clearpage

\appendix

\section[]{The CO Index in Stars}

\begin{figure*}
\vspace{21cm}
\includegraphics{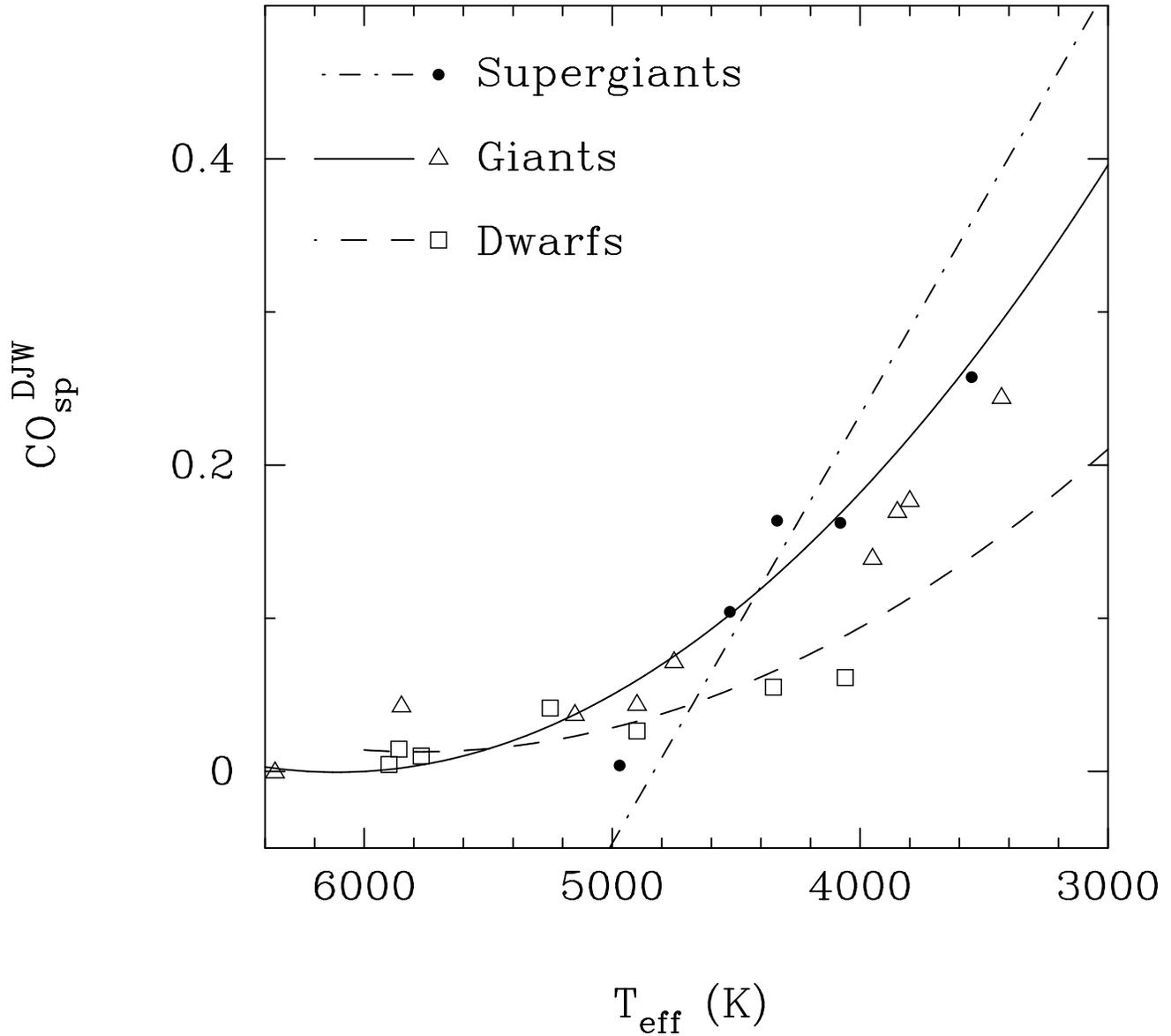}
\caption{Observed variation of the CO spectroscopic index CO$_{\rm sp}^{\rm
DJW}$ (measured over the original wavelength range proposed by DJW) with
spectral type and luminosity class in nearby field stars. Compare this figure
with the equivalent plot in Appendix~A of DJW, which employed the spectral
atlas of Kleinmann \& Hall (1986). The plotted curves are the relations
between CO$_{\rm sp}^{\rm DJW}$ and $T_{\rm eff}$ determined by DJW for
each luminosity class.}
\label{f:costars}
\end{figure*}

Since many of the studies to date of the strength of CO absorption in
starburst systems have been with resolving powers at least twice that
which we have used here, we felt it was important to verify that we
would not lose any sensitivity to stellar effective temperature or
luminosity class. To this end, we set out to observe a variety of
late-type stars covering nearly the full range of spectral type later
than F5, with the same observational setup as for the observations of
M100.  Observing time constraints, coupled with the difficulty in
finding cool stars with well-defined spectral types which were not so
bright as to saturate the detector, precluded us from going below
$T_{\rm eff}=3400$~K.  Table~\ref{t:costars} gives the observed \COsp\
values for both definitions (equations~\ref{eq:codjw} and \ref{eq:pdw7}),
as well as the spectral classification from Hoffleit (1982), the
derived stellar effective temperature from Schmidt-Kaler (1982), and
the continuum power-law slope $\beta$. Figure~\ref{f:costars} presents
the results for our stellar dataset, along with the relations for
dwarfs, giants, and supergiants empirically-determined by DJW from the
stellar atlas of Kleinmann \& Hall (1986).

\begin{figure*}
\vspace{21cm}
\includegraphics{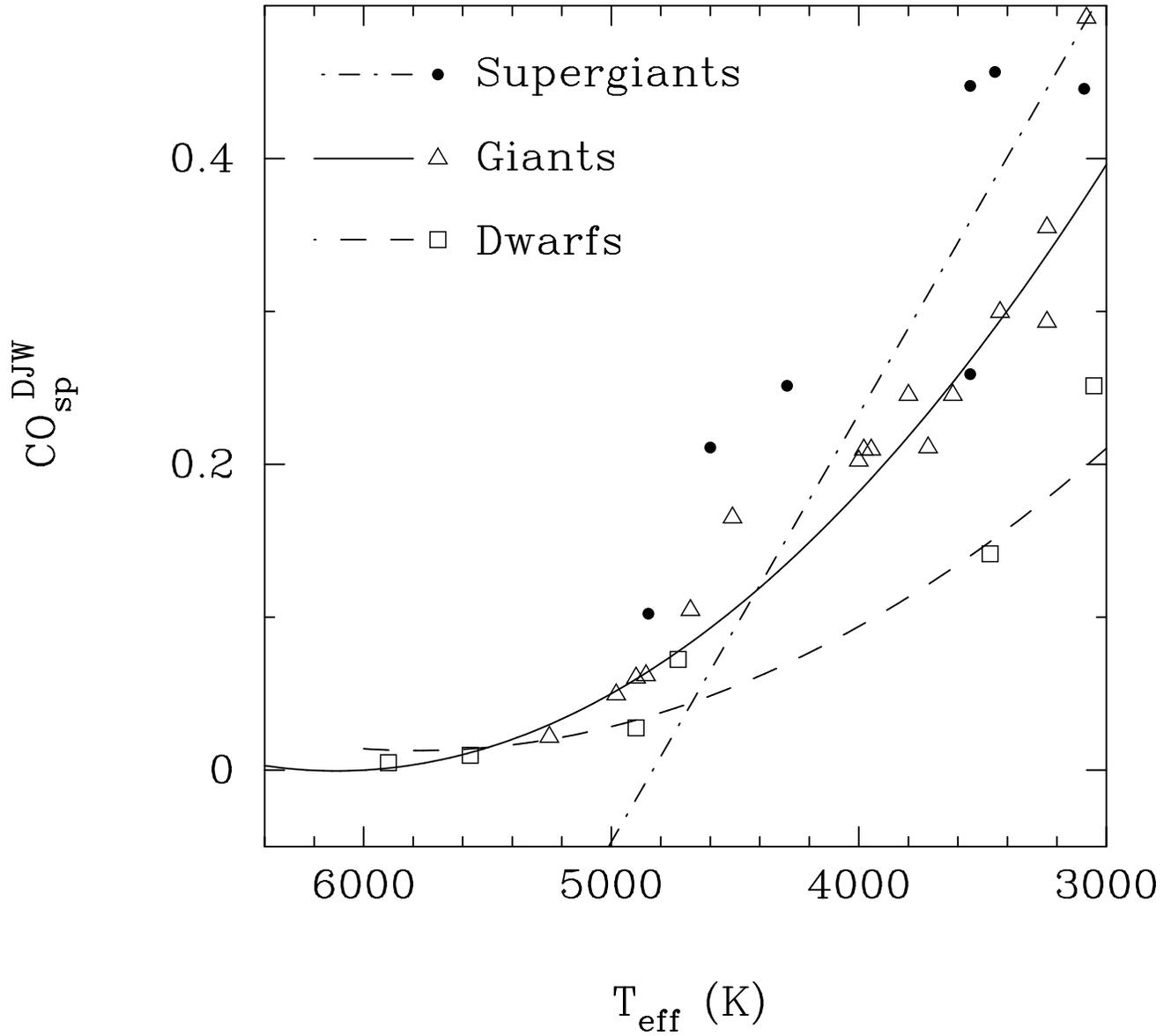}
\caption{As for Figure~\protect{\ref{f:costars}}, but using the $R\sim3000$
stellar spectral atlas of Wallace \& Hinkle (1997).}
\label{f:cowh135}
\end{figure*}

Although our measurements match the expected trends quite well for the
hotter stars, there is considerably more scatter than for the
equivalent plot in DJW (their Fig.~7). In order to assess how much of
this scatter is due to our resolution, and how much may be intrinsic
to the stars themselves, we have made use of the $K$-band stellar
spectral library of Wallace \& Hinkle (1997). We measured CO$_{\rm
sp}^{\rm DJW}$ in all of their luminosity class I, III, and V stars
with $3000 < T_{\rm eff} < 6000$~K that were not also in the Kleinmann
\& Hall (1986) study.  As Figure~\ref{f:cowh135} shows, even at higher
resolution ($R\sim3000$) there is indeed somewhat more scatter in CO
spectroscopic index with type and luminosity class than one might
conclude from the analysis of DJW, particularly amongst the bright
supergiants (a similarly enhanced degree of scatter amongst the
supergiants is exhibited by the independent sample of F\"{o}rster
Schreiber 2000). While this does provide some reassurance that lower
resolution does not significantly compromise our ability to measure
real changes in CO$_{\rm sp}^{\rm DJW}$ with changing stellar
population, the large scatter (up to 0.1~dex) about the mean relation
defined for supergiants by DJW ought to be of concern.  The
supergiants, when present, will dominate CO$_{\rm sp}^{\rm DJW}$, and
the predictions of the models assume a much tighter relation between
CO$_{\rm sp}^{\rm DJW}$ and $T_{\rm eff}$ like the one found by DJW.

\begin{figure*}
\vspace{21cm}
\includegraphics{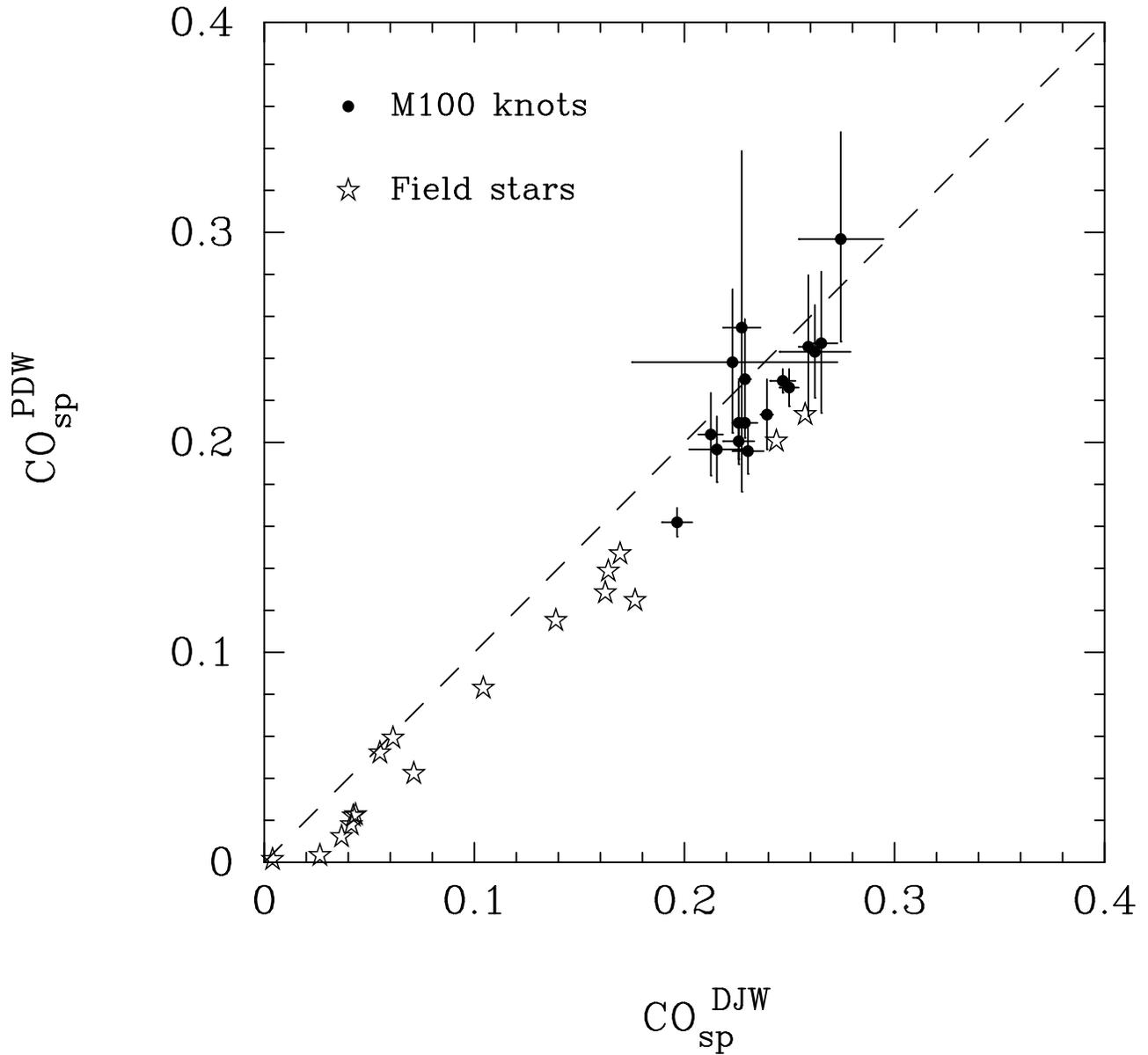}
\caption{Comparison of the CO spectroscopic indices in nearby field
stars and in the M100 knot spectra. CO$_{\rm sp}^{\rm DJW}$ are
the direct integrated measurements between rest wavelengths of
2.31--2.40~$\mu$m, while CO$_{\rm sp}^{\rm PDW}$ are measurements
over the rest wavelength range 2.293--2.320~$\mu$m which are then
converted to the equivalent full-range value. The dashed line
indicates what would be a 1:1 correspondence.}
\label{f:cotest}
\end{figure*}

One aspect that may be resolution-dependent is the conversion from CO
equivalent width over the narrow wavelength interval used by PDW, and
the equivalent CO spectroscopic index (equation~\ref{eq:pdw7}).  Lowering the
resolution will tend to broaden the CO bandhead, and the wavelength
limits may no longer encompass the same extent of the line
profile. The much larger wavelength interval used by DJW will be less
susceptible to this effect. Indeed, when we plot \COsp\ using both
definitions for our own observations of field stars, as well as all the
M100 knots (Figure~\ref{f:cotest}), it is apparent that CO$_{\rm
sp}^{\rm DJW}$ is consistently less than CO$_{\rm sp}^{\rm PDW}$ by
$\sim0.02$~dex when equation~\ref{eq:pdw7} is used with $\eta=16.6$.
Hill et al. (1999) found a similar tendency to underestimate
CO$_{\rm sp}^{\rm PDW}$ in their $R\sim400$ spectra. We can compensate
for this by using $\eta=16.9$ in equation~\ref{eq:pdw7} to restore
good agreement between the two indices.

\begin{table}
 \caption{Stellar Parameters and CO Spectroscopic Indices}
 \label{t:costars}
 \begin{tabular}{clccrr}
\hline
Star    &  Type  &  $\beta$  &  $T_{\rm eff}$  & CO$_{\rm sp}^{\rm DJW}$ &
 CO$_{\rm sp}^{\rm PDW}$ \\
\hline
BS~7503 & G1.5~Vb &   -1.5   &      5900       &      0.00     &  -0.01 \\
BS~6697 & G2~V   &    -1.4   &      5860       &      0.01     &  -0.01 \\
BS~6538 & G5~V   &    -1.4   &      5770       &      0.01     &  -0.01 \\
BS~6301 & K0~V   &    -1.3   &      5250       &      0.04     &   0.02 \\
BS~6806 & K2~V   &    -1.6   &      4900       &      0.03     &   0.00 \\
BS~8085 & K5~V   &    -1.2   &      4350       &      0.06     &   0.05 \\
BS~8086 & K7~V   &    -1.1   &      4060       &      0.06     &   0.06 \\
BS~6531 & F6~III &    -1.7   &      6360       &      0.00     &  -0.02 \\
BS~6466 & G0~III &    -1.6   &      5850       &      0.04     &   0.02 \\
BS~6239 & G5~III &    -1.5   &      5150       &      0.04     &   0.01 \\
BS~6287 & G8~III &    -1.9   &      4900       &      0.04     &   0.02 \\
BS~6307 & K0~III &    -1.4   &      4750       &      0.07     &   0.04 \\
BS~6358 & K5~III &    -2.0   &      3950       &      0.14     &   0.12 \\
BS~6159 & K7~III &    -2.3   &      3850       &      0.17     &   0.15 \\
BS~7244 & M0~III &    -2.4   &      3800       &      0.18     &   0.12 \\
BS~6834 & M4~IIIab &  -2.1   &      3430       &      0.24     &   0.20 \\
BS~8752 & G4~0   &    -0.5   &      4970       &      0.00     &   0.00 \\
BS~6713 & K0.5~IIb &  -2.2   &      4530       &      0.10     &   0.08 \\
BS~6498 & K2~II &     -2.1   &      4340       &      0.16     &   0.14 \\
BS~2615 & K3~Ib &     -1.8   &      4080       &      0.16     &   0.13 \\
BS~6693 & M1~Ib &     -1.6   &      3550       &      0.26     &   0.21 \\
\hline
\end{tabular}
\end{table}

\bsp

\label{lastpage}


\newpage
\begin{thebibliography}{99}

\bibitem{aah} Alonso-Herrero A., Ryder S. D., Knapen J. H., 2001, MNRAS,
              in press (astro-ph/0010522)
\bibitem{nic} B\"{o}ker T., Calzetti D., Sparks W., Axon D., Bergeron L.E.,
              Bushouse H., Colina L., Daou D., Gilmore D., Holfeltz S.,
              MacKenty J., Mazzuca L., Monroe B., Najita J., Noll K., Nota A.,
              Ritchie C., Schultz A., Sosey M., Storrs A., Suchkov A., 1999,
              ApJS, 124, 95
\bibitem{ccm} Cardelli J. A., Clayton G. C., Mathis J. S., 1989, ApJ, 345, 245
\bibitem{d98} Devost D., Origlia L., 1998, in Friedli D., Edmunds M., Robert
              C., Drissen L., eds., ASP Conf. Ser. Vol. 147, Abundance
              Profiles: Diagnostic Tools for Galaxy History. Astron. Soc.
              Pac., San~Francisco, p. 201
\bibitem{djw} Doyon R., Joseph R. D., Wright G. S., 1994a, ApJ, 421, 101 (DJW)
\bibitem{dwj} Doyon R., Wright G. S., Joseph R. D., 1994b, ApJ, 421, 115
\bibitem{nfs} F\"{o}rster Schreiber N. M., 2000, astro-ph/0007324
\bibitem{h99} Hill T. L., Heisler C. A., Sutherland R., Hunstead R. W., 1999,
              AJ, 117, 111 
\bibitem{ybs} Hoffleit D., 1982, Bright Star Catalogue (4th edtn.). Yale
              Univ. Observatory, New Haven, CT
\bibitem{kh}  Kleinmann S. G., Hall D. N. B., 1986, ApJS, 62, 501
\bibitem{95a} Knapen J. H., Beckman J. E., Shlosman I., Peletier R. F.,
              Heller C. H., de~Jong R. S., 1995a, ApJ, 443, L73
\bibitem{95b} Knapen J. H., Beckman J. E., Heller C. H., Shlosman I.,
              de~Jong R. S., 1995b, ApJ, 454, 623
\bibitem{k93} Knapen J. H., Cepa J, Beckman J. E., Soledad del Rio M.,
              Pedlar A., 1993, ApJ, 416, 563 
\bibitem{k00} Knapen J. H., Shlosman I., Heller C. H., Rand R. J.,
              Beckman J. E., Rozas M., 2000, ApJ, 528, 219
\bibitem{kot} Kotilainen J. K., Reunanen J., Laine S., Ryder S. D., 2000,
              A\&A, 353, 834
\bibitem{l99} Leitherer C., Schaerer D., Goldader J. D., Gonz\'{a}lez~Delgado
              R. M., Robert C., Foo~Kune D., de~Mello D. F., Devost D.,
              Heckman T., 1999, ApJS, 123, 3 (SB99)
\bibitem{cg4} Mountain C. M., Robertson D. J., Lee T. J., Wade R., 1990,
              Proc. SPIE, 1235, 25
%\bibitem{o95} Oliva E., Origlia L., Kotilainen J. K., Moorwood A. F. M., 1995,
%              A\&A, 301, 55
\bibitem{o99} Origlia L., Goldader J. D., Leitherer C., Schaerer D., Oliva E.,
              1999, ApJ, 514, 96
\bibitem{o00} Origlia L., Oliva E., 2000, A\&A, 357, 61
\bibitem{o89} Osterbrock D. E., 1989, Astrophysics of Gaseous Nebulae and
              Active Galactic Nuclei. University Science Books, Mill Valley, CA
\bibitem{nr}  Press W. H., Teukolsky S. A., Vetterling W. T., Flannery B. P.,
              1992, Numerical Recipes (Second Edition), Cambridge University
              Press, Cambridge
\bibitem{pdw} Puxley P. J., Doyon R., Ward M. J., 1997, ApJ, 476, 120 (PDW)
\bibitem{phm} Puxley P. J., Hawarden T. G., Mountain C. M., 1990, ApJ, 364, 77
\bibitem{prm} Puxley P. J., Ramsay~Howat S. K., Mountain C. M., 2000, ApJ, 529,
              224
\bibitem{reu} Reunanen J., Kotilainen J. K., Laine S., Ryder S. D., 2000,
              ApJ, 529, 853
\bibitem{r99} Ryder S. D., Knapen J. H., 1999, MNRAS, 302, L7 (Paper~I)
\bibitem{lb}  Schmidt-Kaler T., 1982, in Schiaffers~K., Voight H.~H., eds,
              Landolt-B\"{o}rnstein Numerical Data
              and Functional Relationships in Science \& Technology, Vol.~2:
              Astronomy and Astrophysics, Subvolume~b, Stars \& Star Clusters.
              Springer, NY
\bibitem{s96} Skillman E. D., Kennicutt R. C., Shields G. A., Zaritsky D.,
              1996, ApJ, 462, 147
\bibitem{vk97} Vanzi L., Rieke G. H., 1997, ApJ, 479, 694
\bibitem{wsm} Wada K., Sakamoto K., Minezaki T., 1998, ApJ, 494, 236
\bibitem{wh}  Wallace L., Hinkle K., 1997, ApJS, 111, 445

\end{thebibliography}
\end{document}